# Optical performance monitoring at 640Gb/s via slow-light in a silicon nanowire

B. Corcoran[1], C. Monat[1], M.Pelusi[1], C. Grillet[1], T. P. White[2], L. O'Faolain[2],
T. F. Krauss[2], B. J. Eggleton[1] and David J. Moss[1]

[1] *Institute for Photonics and Optical Sciences (IPOS), Centre for Ultra-high Bandwidth Devices for Optical Systems (CUDOS), School of Physics, University of Sydney, New South Wales 2006 Australia*
[2] *School of Physics and Astronomy, University of St Andrews, St Andrews, Fife, KY16 9SS, UK*
*dmoss@physics.usyd.edu.au*

**Abstract:** We demonstrate optical performance monitoring of in-band optical signal to noise ratio (OSNR) and residual dispersion, at bit rates of 40Gb/s, 160Gb/s and 640Gb/s, using slow-light enhanced optical third harmonic generation (THG) in a compact (80µm) dispersion engineered 2D silicon photonic crystal waveguide. We show that there is no intrinsic degradation in the enhancement of the signal processing at 640Gb/s relative to that at 40Gb/s, and that this device should operate well above 1Tb/s. This work represents a record 16-fold increase in processing speed for a silicon device, and opens the door for slow light to play a key role in ultra-high bandwidth telecommunications systems.

## 1. Introduction

Photonic integrated circuits for all-optical signal processing [1-7] still need substantial development to meet the demanding challenges of high energy efficiency (low power operation) simultaneously with ultra-high bandwidth capability [8,9]. The first goal has been traditionally addressed by increasing the waveguide nonlinear parameter [2], $\gamma = \omega n_2 / c A_{eff}$ (where $A_{eff}$ is the effective mode area, $c$ is the speed of light, and $\omega$ is the pump frequency), to as high as 200 $W^{-1} m^{-1}$ in silicon nanowires [3,10], 100 $W^{-1} m^{-1}$ in hybrid organic-silicon slot waveguides [11] and 93.4 $W^{-1}m^{-1}$ in chalcogenide glass nanotapers [12] – 5 orders of magnitude larger than in silica fibers. This is a result of both extremely tight optical mode confinement afforded by their high refractive index (with $A_{eff}$ often < 0.2$\mu m^2$) together with very high material Kerr nonlinearities ($n_2$ = 100's of times larger than silica glass). A complementary approach is to recycle light within high Q-factor (up to $10^9$) resonant cavities such as pure silica microtoroids [13] and microspheres [14], PhC nanocavities [15] or microrings [16,17]. While this approach has resulted in a wide range of low power CW nonlinear optics [13-19], resonant cavities generally fail to provide a solution to the second goal - that of ultra-high bandwidth operation - since their bandwidths typically range from hundreds of kilohertz [13] to at most a few GHz [15,17].

In contrast, it is possible to design compact slow light photonic crystal (PhC) waveguides [20-24] that can meet the challenges of both ultra-high speed and energy efficient operation [25-27]. Not only does their sub-µm scale optical confinement provide $\gamma$'s comparable with nanowires, but PhC waveguides represent a unique platform where the dispersion of light can be engineered with an unprecedented degree of flexibility. Slight structural modifications of the natural periodicity of the PhC lattice have yielded [25-27] very high (up to $n_g$ = 90) group index waveguides with limited dispersion. As light enters the high group index region of a PhC waveguide, (Figure 1), it slows down and hence is spatially compressed *along* the direction of propagation [24], increasing the optical energy density and hence dramatically enhancing the efficiency of nonlinear optical processes. Slow light has been shown to enhance $\chi^{(3)}$ related nonlinear optical phenomena by a factor $S^2$ for quadratic NL processes [28-32], and $S^3$ for cubic processes [33], where S is the "slow-down" factor given by $S = n_{slow} / n_0$ (where $n_0$ is the background (un-enhanced) refractive index of the medium or waveguide). In particular, the effective nonlinearity parameter, given by $\gamma \times S^2$ in slow PhC waveguides, can be as high as 5000$m^{-1}W^{-1}$ for $S$~10 [29]. Recently, this was exploited to demonstrate slow light enhanced optical third-harmonic generation (THG) [33] and self-phase modulation (SPM) [28,29], in 2D silicon PhC waveguides using optical pulses with only a few watts of peak power and at low repetition rates. Similarly enhanced performance is predicted for four-wave mixing [34] and Raman amplification [35] in this type of structure.

In this paper, we exploit the full potential of slow light PhC waveguides to demonstrate their first application to ultra-high speed all-optical signal processing on a chip. We show that even under extremely high bandwidth operation – approaching 1Tb/s – the effective waveguide nonlinearity is still significantly enhanced by slow light. This low group velocity (*c/40*), along with the low dispersion and moderate linear loss [36] over a very wide bandwidth ( > 1THz) allowed us to achieve optical performance monitoring (OPM) of in-band optical signal to noise ratio (OSNR) and residual dispersion at data rates up to 640Gb/s in 2D silicon PhC waveguides - 16 times faster than the current record of 40Gb/s for a silicon all-optical device [37]. Our device is also much shorter (80µm versus 1cm to 2.5cm ) and operates at much lower power levels ( < 14mW coupled average power versus 100's of mW) than what is typical [37] for silicon nanowires. A key aspect of this work is that it is based on optical THG that arises from the same nonlinearity ($\chi^{(3)}$) that underpins almost all optical signal processing, and so this work opens the door for slow light to play a key role in achieving practical all-optical photonic integrated circuits for future ultra-high bandwidth telecommunications systems.

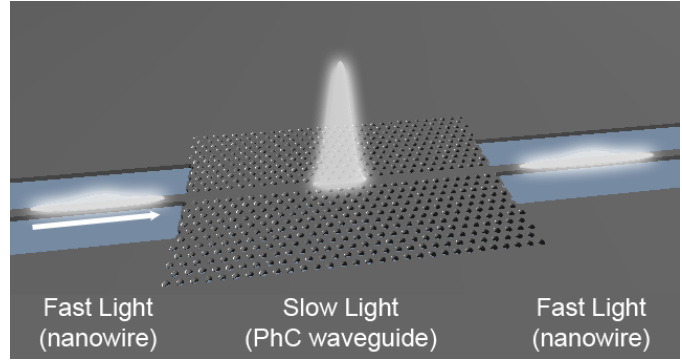

Figure 1: Principle of slow light enhanced nonlinearities in 2D dispersion engineered photonic crystal waveguides.

## 2. Principle of Operation

Optical performance monitoring [38-41] is critical for ultra-high bandwidth optical communications systems where signals become much more sensitive to transmission impairments such as amplified spontaneous emission (ASE) noise, chromatic dispersion, polarization mode dispersion, loss and others. Without monitoring signal quality, a reduction in network performance is only discovered once actual data is lost. However, at very high data rates electronic monitoring methods become impractical. All-optical performance monitors can generate a DC signal that is a direct reflection of the signal quality ("eye" opening, for example), even in the Tbit/s regime; the higher the quality of the data, the higher the average output signal of the optical monitor. All-optical techniques can also provide feedback to signal conditioning devices, such as tunable dispersion compensators [42,43], to allow continuous real-time optimization of signal quality. While various approaches have been taken to monitor the optical performance of networks, such as noise discrimination through measuring polarization state [44], asynchronous sampling [45], radio frequency optical spectrum monitoring [4], and others, an important method that has attracted significant interest is based on a simple nonlinear power transfer scheme [38-41]. Figure 2 illustrates the principle of this approach. Two signals with the same average power but differing degrees of degradation (due to noise, dispersion etc), evidenced by the quality of the "eye" opening, are transformed differently through the device nonlinear transfer function. The high quality pulse train emerges with a greater average power than the noisy signal since the peak to average power ratio is statistically higher. Thus, a simple average power measurement with a slow detector can distinguish between varying degrees of noise or other underlying distortions such as residual dispersion. This approach has been demonstrated using various third-order ($\chi^{(3)}$) nonlinearities to produce the key nonlinear transfer function such as two-photon detection [38,39], cascaded four-wave mixing (FWM) in an optical parametric amplifier (OPA) [41,46,47] or self (SPM) phase modulation [48]. Our use of optical THG – its first for signal processing – is expected to achieve better performance since the associated nonlinear transfer function, i.e. a *cubic* ($\sim [I(\omega)]^3$) function of signal peak power, is steeper than other $\chi^{(3)}$ processes (which vary quadratically with signal peak power) [40].

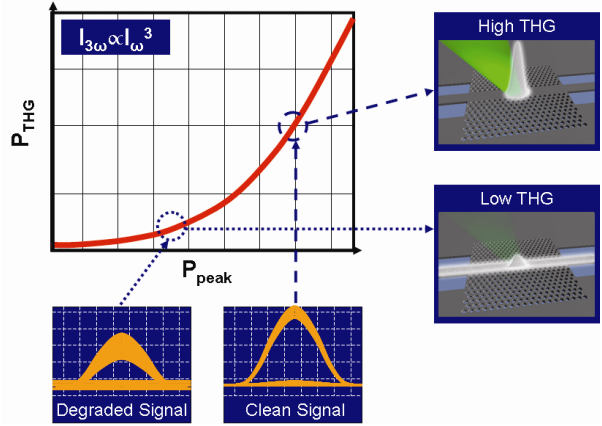

Figure 2 Principle of operation of OPM based on THG. Two signals with the same average power but a different eye diagram quality are converted by the device nonlinear power transfer curve into a bright or faint green light for the undistorted and noisy signal, respectively.

## 3. Device

Figure 1 shows a schematic of the device, a dispersion-engineered slow light PhC waveguide on a suspended membrane, fabricated from a silicon-on-insulator (SOI) wafer and connected to two tapered ridge access nanowires. The 2D PhC structure consists of a triangular lattice of air holes with lattice constant a=414nm and hole radius of 118nm (0.286a) etched into a 220nm thick silicon suspended membrane. A W1 waveguide is introduced by omitting a single row of holes along the ΓK direction to form a linear defect. The total PhC waveguide length is 80µm, and the lattice period of the first and last 10 periods is increased to 444nm parallel to the waveguide to enhance coupling from the nanowire to the slow light mode [49]. The dispersion of the PhC waveguide is engineered so as to provide a high group index region with reduced dispersion by shifting the first two rows of holes adjacent to the guide perpendicular to the direction of propagation [25]. For the waveguide used in this experiment, the first and second rows are shifted 52nm away from and 12nm toward the axis of the waveguide, respectively. Light is coupled in and out of the PhC waveguide via 2mm long ridge access waveguides whose width is tapered from 3µm to 0.7µm over 200µm close to the PhC waveguide. The device was fabricated from a SOITEC silicon-on-insulator wafer by electron-beam lithography (hybrid ZEISS GEMINI 1530/RAITH ELPHY) and reactive ion etching using a $CHF_3/SF_6$ gas mixture. The silica layer under the PhC slab was selectively under-etched using a HF solution to leave the PhC section in a suspended silicon membrane. More details of the procedure are given in [25]. The resulting "flat band slow light" region is shifted away from the edge of the first Brillouin zone, where high loss and high dispersion have traditionally been problematic [21]. Figure 3 shows the (measured) group index versus wavelength for the particular device used in this work, designed to have $n_g$ ~ 38 ± 10% over a ~12nm wavelength range near 1560nm. These parameters were chosen as a compromise between wide bandwidth and high group index [25]. Also shown in Figure 3 is the optical spectrum of the highest bandwidth signal used in these experiments (640Gb/s), along with its "eye" diagram (inset). The fact that the 3dB bandwidth of the 640Gb/s signal entirely fits within the "flat band slow light" window clearly implies that slow light in this structure should enhance nonlinear optics equally well at both low, and extremely high, bit rates.

The typical second-order dispersion ($\beta_2$) in the dispersion-engineered slow PhC waveguides reported here is ~1000 $ps^2/m$ [28], which yields a dispersion length ($L_D$) of ~250µm for a pulsewidth of 520fs (the shortest pulses considered here). This is considerably longer than the 80µm PhC waveguide length and so we expect the transmitted signal to be largely unaffected, other than the (modest) linear propagation loss across the PhC, which is ~ 0.4dB (~ 50dB/cm) [36]. We note that this device was not optimized in terms of coupling losses, and estimate that ~10% of the light launched to the chip is actually coupled into the PhC waveguide. The measured total loss of ~ 24dB was dominated by end-facet coupling.

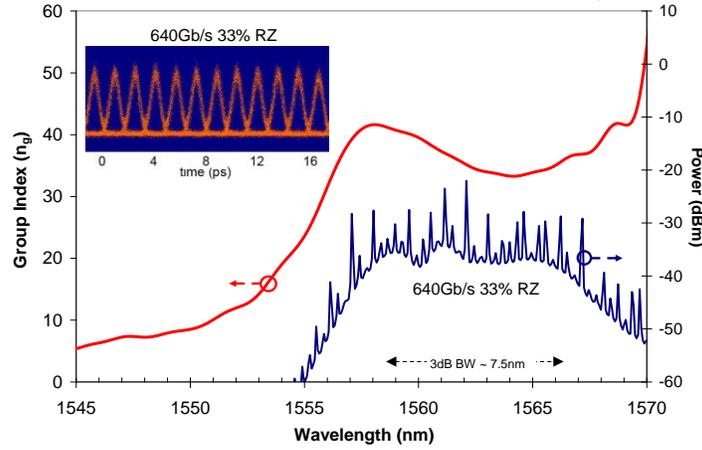

Figure 3: Group index versus wavelength for the silicon engineered PhC waveguide used in this work, having a group index $n_g$~38, nearly constant (±10%) over a wide (~12nm) spectral region. Also shown is the optical spectrum of the 640Gb/s PRBS data stream used in the experiments, showing that the slow light bandwidth is wide enough to accommodate the data bandwidth. Inset: eye diagram of 640Gb/s 33% RZ PRBS modulated signal, as measured with an all-optical sampling oscilloscope.

## 4. Experiment

To examine the dependence of THG efficiency on signal bit rate, we coupled optical signals with different bit rates into the device and measured the average power of the THG – visible green light at 520nm, emitted out-of-plane of the PhC waveguide and imaged onto a sensitive detector. The input signals were spectrally tuned to near the centre of the slow light region and modulated with pseudo-random data at bit rates of 40Gb/s, 160Gb/s and 640Gb/s. Figure 4 shows the experimental setup for monitoring either residual dispersion or OSNR. The data source was a tunable 40GHz modelocked laser ($\lambda$= 1560.2nm, 1.1ps, 4% duty cycle), followed by a pulse compressor comprised of 500m HNLF ($\gamma$=30 W.km$^{-1}$) and a 5nm/8nm spectral filter (N.B. this pulse compression step is skipped for 40GBit/s data) to achieve variable pulsewidths (see Table 1). This was then modulated with a pseudo-random bit sequence ($2^{31}$-1 PRBS) and multiplexed up to bit rates of 160Gb/s and 640Gbit/s using a time interleaver ($2^7$-1 pattern length), yielding a return-to-zero (RZ) modulated signal. The signal then underwent dispersion control using a Finisar WaveShaper [50]. For the OSNR measurements, the signals were combined with a variable amount of amplified spontaneous emission (ASE) noise from an erbium doped fiber amplifier (EDFA), filtered by a 5nm (for 40Gbit/s and 160Gbit/s data) or 8nm (640Git/s data), equal to the signal bandwidth at the different bit rates, to ensure that the noise was in-band. The ratio of signal to noise was controlled by adjusting air-gap variable fibre attenuators before mixing signal and noise. The polarization of the combined polarized signal and noise was controlled with a fibre polarization controller and in-line polarizer such that the TE mode of the PhC waveguide is excited. Coupling was achieved via a lensed fibre with a 2.5µm focal spot, to the 3 µm wide silicon access ridge waveguide. The third harmonic light (around 520nm) emitted out of the plane of the PhC waveguide by the photonic crystal, was collected by a 20x, 0.42 N.A. long working distance objective and imaged onto an amplified silicon photodiode, with a specified conversion gain at 520nm of ~0.325 V/pW. The average power of the THG signal was monitored as the signal impairment (alternatively dispersion or OSNR) was varied, at different bit-rate signals, while keeping the average coupled power fixed (< 14mW, Table 1).

The THG was collected and detected from the entire length of the PhC waveguide, although the majority of the THG was emitted from the first half of the slow PhC waveguide since we observed the same exponential decay in the THG as we reported earlier [33]. In principle this implies that the length of the device would only need to be ~ 40 μm to achieve the same efficiency as we report here. Note also that the device is polarization sensitive, in that the waveguide is engineered for enabling slow light propagation and in turn, THG, only for the TE mode. Hence, this device is sensitive only to the TE polarized near-infrared input.

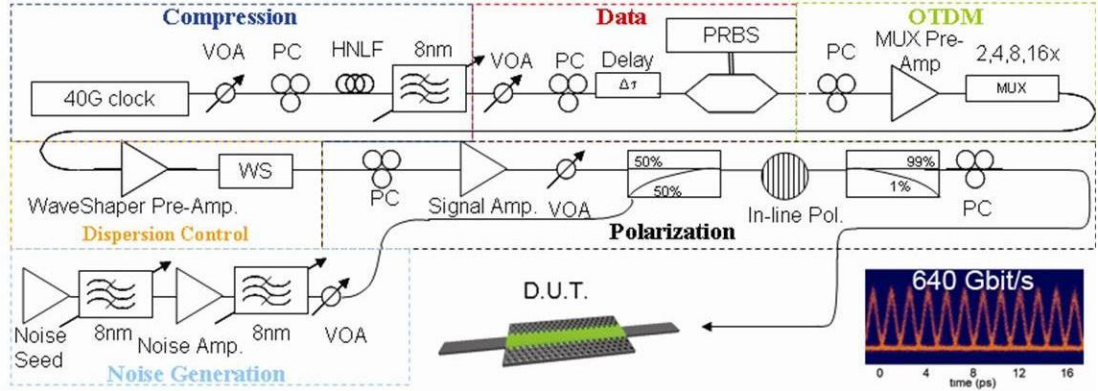

Figure 4 Experimental setup used to measure both THG as well as perform OPM using THG of a 640GBit/s data signal. For the 40 and 160GBit/s signals, the 8nm filters are replaced with 5nm filters, and launch power varied into the compression set up. Inset: Optically sampled eye diagram of 640GBit/s data signal.

## Table I
### Optical Pulse Parameters.
(Powers correspond to the average power coupled into the waveguide)

| **OSNR Monitoring** Bit Rate (Gb/s) | Pulsewidth (ps) | Duty Cycle | | Power |
|---|---|---|---|---|
| 40 | 2 | 8% | | 8mW |
| 160 | 1.3 | 14% | | 8mW |
| 640 | 0.52 | 33% | | 14mW |
| **Dispersion** Bit Rate (Gb/s) | Pulsewidth (ps) | Bandwidth (nm) | Duty Cycle | Power |
| 40 | 1.1 | 2 | 4% | 7mW |
| 160 | 0.85 | 4 | 14% | 10mW |
| 640 | 0.52 | 7.5 | 33% | 12mW |

**5. Results**

Figure 5a shows the cubic dependence of the third harmonic power as a function of peak near-infrared coupled power at each bit rate. The apparent increase in the THG signal at 640 Gbit/s compared to 160 and 40 Gbit/s is due to the different duty cycles used. When this effect is normalised out (Figure 5b), the curves almost coincide, confirming that there is no intrinsic "roll-off" in efficiency of the device at higher bit rates and that the bandwidth of this device is large enough to accommodate most of the 640Gb/s signal. Figure 5a also shows that if the 640Gb/s signal is tuned (λ=1542nm) away from the slow light (high $n_g$) region, no TH signal is detected above the dark noise, clearly indicating that slow light plays a critical role in enhancing the THG efficiency. Note that at the power levels (< 14mW coupled average power) used in these experiments, we do not observe saturation due to two-photon absorption and the corresponding free carrier generation - a well known effect in silicon.

Figure 6 summarizes the results of the residual dispersion monitoring experiments at bit rates of 40Gb/s, 160Gb/s, and 640Gb/s, and clearly shows that the THG efficiency strongly varies with the residual dispersion of the incident pulse-train, which is reflected in the degree of "eye" closure of the modulated PRBS signal. The sensitivity of the THG to residual dispersion increases dramatically with bit rate (ie., the THG decreases for much smaller values of residual dispersion), as expected [41]. The associated dynamic range, however, decreases with increasing bit rate, primarily because the duty cycle for the 640Gb/s signal is higher than at lower bit rates as discussed further below. The secondary peak visible in the 640Gb/s trace may be due to temporal Talbot interference between neighboring pulses [48].

Figure 7 shows the relative average THG power (in dB) as a function of input OSNR (for a fixed average coupled power) at all three bit rates, and clearly indicates that the THG is a strong function of input OSNR. The results agree well with calculations which estimate the change in emitted THG power with varied OSNR, using the probability density distribution of the signal and noise power, passed through the cubic transfer function [41]. The only parameter changed for the three calculated curves was signal duty cycle, to correspond to with the duty cycles used in the experiments at different bit rates (8%, 14% and 33% - see Table I). The slight disagreement at 40Gb/s and 160Gb/s is within the experimental uncertainty. The reduction in dynamic range we observe as the bit rate is increased is a generic characteristic of the nonlinear power transfer function approach to OPM, which becomes less sensitive at higher duty cycles [41]. In order to isolate this effect we plot the experimental curves for a 40Gb/s 33% RZ signal, which shows good agreement with the 33% theoretical curve. The corresponding curve for 640Gb/s deviates from this by about 1dB at the lower limit of the OSNR range, and is a result of the initial (noise free) 640Gb/s pulse shape being not quite as ideal as in the experiments summarized in Figure 5, rather than a limitation or "roll off" in the intrinsic device bandwidth.

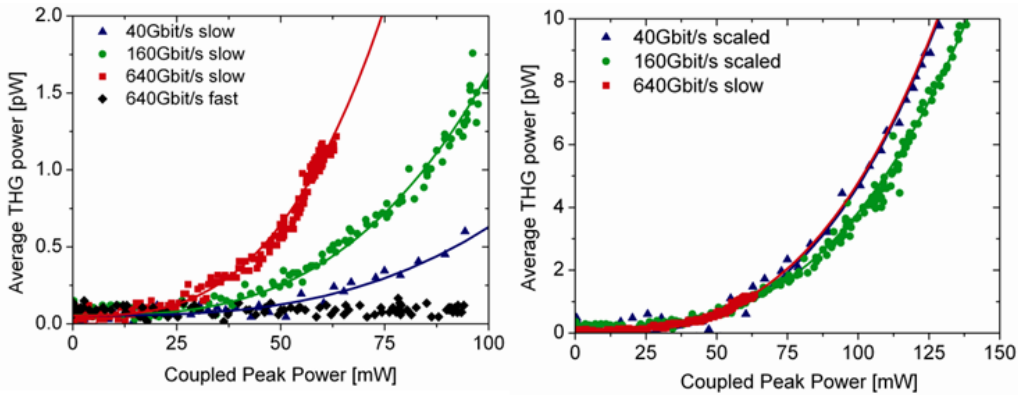

Figure 5. Average THG power versus near-infrared coupled peak power for "clean" (non-degraded) signals at bit rates of 40Gbit/s, 160Gbit/s and 640Gbit/s along with cubic fits to the data. Left: a). The signal for all measurements was tuned to within the slow light region, except black square data which were taken at 640Gbit/s tuned to outside of the SL regime (1542nm). Right b). The normalized curves shown here for 40 and 160Gbit/s were obtained by scaling the raw data by the ratio of signal duty cycle to 640Gbit/s duty cycle (i.e. 4%/33% scaling for 40Gbit/s, 14%/33% scaling for 160Gbit/s).

## 6. Discussion

In addition to representing the first application of slow light to high speed signal processing, this work achieves a leap in processing speed of 16 times over previous reports of all-optical signal processing in a silicon device. It is also the highest bit rate for which OPM has been reported using any technique. In addition to these, it is the first application of optical THG to any form of signal processing. The associated emission of visible light provides a powerful indicator that can be readily measured using sensitive detectors such as avalanche photodiodes and possibly integrated devices. The intrinsic cubic nonlinearity of the THG process is unique for $\chi^{(3)}$ processes – all of the rest (self and cross-phase modulation, four wave mixing etc.) depending either linearly or quadratically on pump (or signal) power. This represents a significant advantage for THG in terms of nonlinear signal processing, which relies on the degree of nonlinearity of the underlying process. This is reflected in the substantial improvement of our results over previous reports of optical performance monitoring, based on cascaded four-wave mixing (CFWM) [47] at 10Gb/s. Our results at 160Gb/s (where we have a similar duty cycle to [47], of ~14%) show a significantly higher dynamic range (6dB) than in [47] (1.5dB). The operating power of our device can in principle be significantly reduced, potentially to sub-milliwatt levels, by employing inverse-tapers to minimize coupling loss. Improving the THG efficiency should also be possible with a greater understanding of the underlying THG process, including the role of phasematching in this system, where the absorption length of the TH light is on the order of 1μm. This is investigated in more detail elsewhere [52].

The nonlinear losses in the regime the device was operated in these experiments were negligible, while the linear propagation loss (Section 3) was ~ 0.4dB. Further, the effects of dispersion, even for the shortest pulses studied here (520fs), were restricted given that the dispersion length is ~ 250μm (ie., 3x longer than the 80μm PhC length). Considering the exponential decay of the TH light along the waveguide, even greater performance could be achieved using a shorter (40μm) device that could be readily used "in-line" where the signal is transmitted through the device, rather than being "tapped off" or sampled. Hence, this would advantageously provide a non destructive method for monitoring the quality of the signal without degrading it.

The ultimate speed limit of this device is not determined by the underlying physical process, since $\chi^{(3)}$ is a virtual process with an intrinsic speed on the order of a femtosecond or less [53], but rather only by the bandwidth of the slow light region. For our device this is ~ 12nm, indicating that it should be capable of processing signals at well above terabit per second speeds. Note that in general there is an inverse relationship between bandwidth and group index, to the extent that it is useful to define a figure of merit given by the group index - bandwidth product, $FOM = n_g \Delta\omega/\omega$. For the approach to slow light engineering that we have used in this work [25], this FOM is ~ 0.3, and so larger bandwidths than the 12nm reported here can be obtained but with a corresponding reduction in slow light enhancement. A higher FOM – in excess of 0.4 – may be achieved in the future, which would allow the monitoring of even higher speed Tb/s pulses with the same slow light enhancement factor as reported here. These speeds could be comparable with the fastest analogue signal processing reported to date in integrated devices, at several THz [4,6].

Finally, while OPM of high speed signals is a relatively specific function, this work is based on the same third-order optical nonlinearity ($\chi^{(3)}$) as most forms of high-speed all-optical signal processing. Further, the application of these devices to new coherent modulation formats [51] can be readily achieved with the inclusion of well known integrated devices such as Mach-Zehnder interferometers. The result is that the door has now been opened for slow light to play a key role in achieving ultra-high bandwidth all-optical photonic integrated circuits for future optical telecommunications systems, potentially to well beyond 1Tb/s.

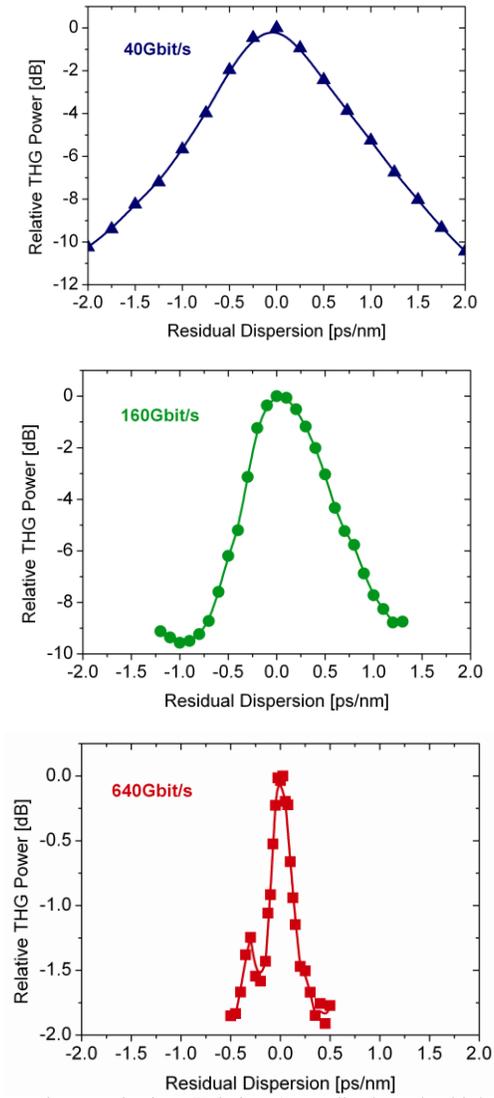

Figure 6. Residual dispersion monitoring: Relative (normalised to the highest monitor reading) THG average power versus residual dispersion setting for fixed coupled power at 40Gb/s (top) 160Gb/s (middle) and 640Gb/s (bottom), respectively.

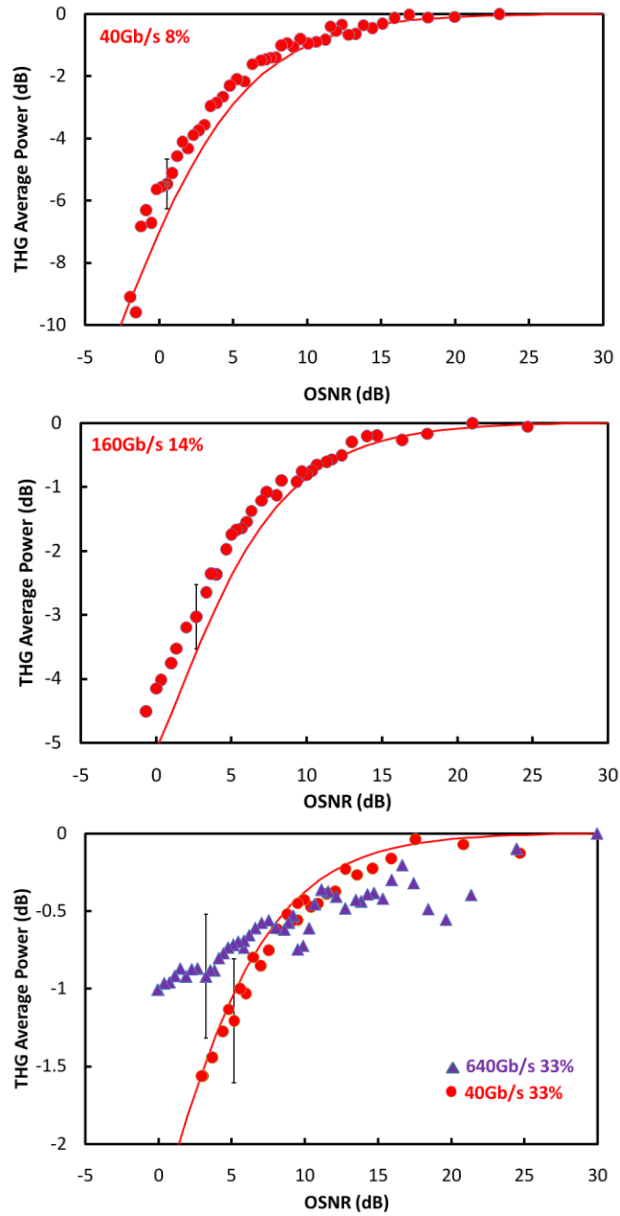

Figure 7. OSNR monitoring: Relative (normalised to the high OSNR value) THG average power versus input OSNR for fixed coupled power (from top to bottom) at 40Gb/s (top), 160Gb/s (middle) and 640Gb/s (bottom), respectively. Solid lines indicate theoretical curves associated with the three duty cycles used in the experiments. The bottom graph also contains experimental and theoretical results for 40Gbs at 33% duty cycle.

**7. Conclusion**

We demonstrate optical performance monitoring of in-band optical signal to noise ratio (OSNR) and residual dispersion, at bit rates of 40Gb/s, 160Gb/s and 640Gb/s, using slow-light enhanced optical third harmonic generation (THG) in a dispersion engineered 2D silicon photonic crystal waveguide. We show that there is no intrinsic degradation in the enhancement of the signal processing at 640Gb/s relative to that at 40Gb/s, and that this device should operate well above 1Tb/s. This work represents a record 16-fold increase in processing speed for a silicon device, and opens the door for slow light to play a key role in ultra-high bandwidth telecommunications systems.

**Acknowledgments**
The authors would like to thank the EU-FP6 Network of Excellence ePIXnet, for the fabrication, which was carried out in the framework of the ePIXnet nanostructuring platform for photonic integration (www.nanophotonics.eu).We would also like to acknowledge support from the Australian Research Council (ARC) including the ARC Centre of Excellence program.